%% file: main.tex
\documentclass[12pt]{article}
\input{_preamble}

\title{\titlepaper}

\begin{document}
\maketitle

\begin{abstract}
  Marginal structural models are a popular method for estimating causal
  effects in the presence of time-varying exposures. In spite of their
  popularity, no scalable non-parametric estimator exist for marginal
  structural models with multi-valued and time-varying treatments.
  In this paper, we use machine learning together with recent
  developments in semiparametric efficiency theory for longitudinal
  studies to propose such an estimator. The proposed estimator is
  based on a study of the non-parametric identifying functional,
  including first order von-Mises expansions as well as the efficient
  influence function and the efficiency bound. We show conditions
  under which the proposed estimator is efficient, asymptotically
  normal, and sequentially doubly robust in the sense that it is
  consistent if, for each time point, either the outcome or the
  treatment mechanism is consistently estimated.  We perform a
  simulation study to illustrate the properties of the estimators, and
  present the results of our motivating study on a COVID-19 dataset studying the impact of mobility on the cumulative number of observed cases.
\end{abstract}

\section{Introduction}

In a counterfactual framework for causal inference with time-varying
exposures, causal effects are often defined as contrasts between the
marginal distributions of the hypothetical outcomes $\{Y(\bar a):\bar a\}$ that would have
been observed if, possibly contrary to fact, the time-varying exposure
had been set to a fixed vector value $\bar
a=(a_1,\ldots,a_\tau)$. When the number of time points $\tau$ or the
cardinality of the support of treatment increases (e.g., when the
treatment is numerical), the curse of dimensionality makes it
challenging to estimate the distribution of $Y(\bar a)$. Marginal
structural models \citep{robins1997marginal}, which in their most popular form rely on an assumption that the
expectation of $\E[Y(\bar a)]$ follows a (semi)parametric model as a
function of $\bar a$, have become a popular approach to address this
problem.
Multiple estimators including outcome regression, inverse probability
weighted, and doubly robust estimators have been proposed for marginal
structural models \citep[e.g.,][]{robins2000marginal, Hernan00,
  Robinsetal00, Bang05, saarela2015bayesian}, with the inverse
probability weighting approach of \cite{Robinsetal00} being the most
commonly used in practice. While marginal structural models have
spurred much progress, a few important methodological limitations
remain unaddressed.

First, in most real life studies investigators will not know the
correct parameterization of the model for $\E[Y(\bar a)]$, even if it
existed. While methods have been developed for model selection for
marginal structural models \citep{vanderLaan&Dudoit03,
  brookhart2006semiparametric, platt2013information,
  baba2017criterion, sall2019test}, they lack theoretical a foundation
for post-selection inference with guaranteed frequentist operating
characteristics (e.g., nominal coverage of confidence intervals, error
control for hypothesis tests, etc.). Applied research then must
proceed using one of three unsatisfactory options: (i) using an
incorrect but pre-specified marginal structural model, (ii) using
model selection but incorrectly quantifying statistical uncertainty as
if the model was pre-specified, or (iii) ad-hoc re-sampling
post-selection inference methods such as the bootstrap for which there
are not any general theoretical guarantees.

Second, the consistency of inverse probability weighted estimators, the most popular approach in practice,
relies on the correct specification of the models for the weights. If
these models are parametric and correctly specified, the delta method
or the bootstrap may provide a basis for valid statistical
inference. However, as with the marginal structural model itself, the
correct parameterization for the models for the exposure probabilities
will not be known a priori, and model selection is
necessary. Moreover, if the dimension of the covariates is moderate to
large, flexible regression methods from the machine and statistical
learning literature are often necessary to achieve an appropriate fit
\cite[e.g., ][]{lee2010improving, diaz2011super, gruber2015ensemble,
  bentley2018impact}. However, inverse probability weighted estimators
using flexible regression or model selection for the weights also lack
a general theoretical foundation for statistical inference with
guaranteed operating characteristics. Similar issues affect estimators
based on outcome regression and the doubly robust semiparametric
estimators of \cite{Bang05}.

These two problems result in biased analyses that produce confidence
sets with incorrect coverage and null hypothesis tests with incorrect
type I and II error \citep{mortimer2005application,
  lefebvre2008impact, lipkovich2012challenges}.

While approaches to address misspecification of the marginal
structural model are available, they have their own
limitations. Specifically, \cite{Neugebauer2007419} proposed to
address this problem by defining the estimand as the projection of
$\E[Y(\bar a)]$ onto the posited model, rather than assuming that the
model is correctly specified. This approach requires that that the
model is useful and parsimonious rather than correct, and therefore
explicitly captures the idea that models must be viewed as
approximations \citep[see e.g.,][]{box1979robustness, buja2019models2,
  davison2019comment}. 
The approach of defining the target parameter as a projection onto a
model has a long history and is widely used in statistics
\citep[e.g.,][]{huber1967behavior, beran1977minimum, white1980using,
  wasserman2006all,vansteelandt2022assumption} as well as in causal
inference
\citep[e.g.,][]{van2006statistical,cuellar2020non,kennedy2021semiparametric}.

Multiple non-parametric efficient estimators leveraging flexible
regression have been proposed for the projection parameter of
\cite{Neugebauer2007419} in the case of categorical exposures, but
they are not scalable in the number of categories. Furthermore, none
of these estimators can be used with continuous time-varying
exposures. Existing estimators can only be used with categorical
exposures taking on a few values because they require creating
datasets that pool over regimes $\bar a$
\citep[e.g.,][]{rosenblum2010targeted, petersen2014targeted}, or else
they require to estimate sequential regressions separately for each
possible regime $\bar a$ \citep[e.g.,][]{schnitzer2014modeling}.

In this manuscript we develop non-parametric estimators for marginal
structural models with the following properties. First, the estimators
can accommodate time-varying treatments of any kind, continuous,
binary, categorical, etc. Second, the estimators can leverage flexible
regression techniques from the machine and statistical learning
literatures for improved consistency. Third, we prove that the
estimators are sequentially doubly robust in the sense that they are
consistent under an assumption that, at each time point, one of two
nuisance parameters is consistently estimated. Fourth, the estimators
are root-n consistent, asymptotically normal, and efficient under
consistent estimation of all nuisance parameters at
slower-than-parametric rates (e.g., $n^{1/4}$).

Our estimators are rooted in semiparametric efficiency theory
\citep[e.g.,][]{mises1947asymptotic, vanderVaart98,
  robins2009quadratic, Bickel97, Robins00,vanderLaan2003, Bang05}, in
sequential doubly robust estimators for the g-computation formula
\citep[e.g.,][]{rubin2007doubly, van2012targeted,
  rotnitzky2012improved, luedtke2017sequential, rotnitzky2017multiply,
  molina2017multiple,diaz2021nonparametricmtp, diaz2022causalcomp},
and in recently proposed methods for obtaining estimators with doubly
robust asymptotic distributions \citep[e.g.,][]{benkeser2016doubly,
  diaz2017doubly, diaz2019statistical}. Central to this theory and to
our proposal is the study of the so-called von-Mises expansion
\citep{mises1947asymptotic} and the associated canonical gradient and
second-order term, which characterize the efficiency bound of the
target functional and allow the development of estimators under slow
convergence rates for the nuisance parameters involved
\citep{robins2009quadratic}. We use cross-fitting to obtain root-n
convergence of our estimators while avoiding entropy conditions that
may be violated by data adaptive estimators of the nuisance parameters
\citep{zheng2011cross, chernozhukov2018double}.

\section{Notation and introduction of the problem}
Let $Z_1,\ldots, Z_n$ denote a sample of i.i.d. observations with
$Z=(L_1, A_1, L_2, A_2, \ldots, L_\tau, A_\tau, Y)\sim \P$, where
$L_t$ denotes time-varying covariates, $A_t$ denotes a general vector of
 exposure or treatment variables, and $Y=L_{\tau+1}$ denotes
an outcome such as survival at the end of study follow-up. We let
$\P f = \int f(z)\dd \P(z)$ for a given function $f(z)$. We use $\Pn$
to denote the empirical distribution of $Z_1,\ldots\,Z_n$, and assume
$\P$ is an element of the nonparametric statistical model defined as
all continuous densities on $Z$ with respect to a dominating measure
$\nu$. We let $\E$ denote the expectation with respect to $\P$, i.e.,
$\E\{f(Z)\} = \int f(z)\dd\P(z)$. We also let $||f||^2$ denote the
$L_2(\P)$ norm $\int f^2(z)\dd\P(z)$. We use
$\bar X_t = (X_1,\ldots, X_t)$ to denote the history of a variable,
and use $H_t = (\bar A_{t-1}, \bar L_t)$ to denote the
history of all variables up until just before $A_t$. For the complete
history of a random variable, we simplify $\bar X_\tau$ as $\bar
X$. We let $\g_t(a_t \mid h_t)$ denote the probability mass or density
function of $A_t$ conditional on $H_t=h_t$. We use calligraphic font
to denote the support of a random variable, e.g., $\mathcal A_t$
denotes the support of $A_t$. We will use $V$ to denote a set of
baseline covariates of interest. By convention, variables with an
index $t\leq 0$ are defined as the null set, expectations conditioning
on a null set are marginal, products of the type
$\prod_{t=k}^{k-1}b_t$ and $\prod_{t=0}^0b_t$ are equal to one, and
sums of the type $\sum_{t=k}^{k-1}b_t$ and $\sum_{t=0}^0b_t$ are equal
to zero. For two vectors $v,u\in\mathbb{R}^d$, we let $u\cdot v$
denote the usual dot product.

We formalize the definition of the causal effects using a
non-parametric structural equation model
\citep{Pearl00}. Specifically, for each time point $t$, we assume the
existence of deterministic functions $f_{L_t}$, $f_{A_t}$, and $f_Y$
such that $L_t=f_{L_t}(A_{t-1}, H_{t-1}, U_{L,t})$,
$A_t=f_{A_t}(H_t, U_{A,t})$, and $Y=f_Y(A_\tau, H_\tau, U_Y)$. Here
$U=(U_{L,t}, U_{A,t}, U_Y:t\in \{1,\ldots,\tau\})$ is a vector of
exogenous variables, with unrestricted joint distribution. Causal
effects can be defined in terms of hypothetical interventions where
the equation $A_t=f_{A_t}(H_t, U_{A,t})$ is removed from the
structural model, and the exposure is assigned as a fixed value
$a_t$. An intervention that sets the exposures up to time $t-1$ to
$\bar a_{t-1}$ generates counterfactual variables
$L_t(\bar a_{t-1}) = f_{L_t}(a_{t-1}, H_{t-1}(\bar a_{t-2}),
U_{L,t})$, where the counterfactual history is defined recursively as
$H_t(\bar a_{t-1}) = (\bar a_{t-1}, \bar L_t(\bar a_{t-1}))$.  An
intervention where all the treatment variables up to $t=\tau$ are
intervened on generates a counterfactual outcome
$Y(\bar a)=f_Y(a_\tau, H_\tau(\bar a_{\tau-1}), U_Y)$. Causal effects
will be defined in terms of the distribution of this counterfactual.

The expectation of $Y(\bar a)$ is identifiable using the g-computation
formula \citep{Robins86} as follows. Define the following assumptions:
\begin{assumption}[Sequential randomization]\label{ass:seqign}
  Assume $U_{A,t}\indep (U_{L,t+1},\ldots,U_{L,\tau+1})\mid H_t$ for
  all $t\in\{1,\ldots,\tau\}$.
\end{assumption}
\begin{assumption}[$\lambda$-positivity of treatment assignment mechanism]\label{ass:pos}
  For a user-given density function $\lambda$ on $(\bar a, v)$, if
  $\lambda(a_t\mid \bar a_{t-1}, v)>0$, then $\P[\g_t(a_t\mid H_t) > 0]>0$ for all
  $t\in\{1,\ldots,\tau\}$.
\end{assumption}
Set $\Q_{\tau+1}= Y$. For $t=\tau,\ldots,1$, recursively define
\begin{equation}
  \Q_t:(\bar a_t, h_t) \mapsto \E\left[\Q_{t+1}(\bar a_{t+1}, H_{t+1})\mid
    A_t=a_t, H_t=h_t\right],\label{eq:defQ}
\end{equation}
and define $\theta(\bar a, v) = \E[\Q_1(\bar a, L_1)\mid V=v]$. Under
Assumptions \ref{ass:seqign} and \ref{ass:pos} we have
$\E[Y(\bar a)\mid V=v]=\theta(\bar a, v)$ whenever
$\lambda(\bar a, v)>0$.

When the interest is to estimate the parameter $\theta(\bar a, v)$
only at a few values $\bar a$, estimation can proceed by sequentially
fitting the regressions in equation (\ref{eq:defQ}). If these
regressions are fitted in correctly pre-specified parametric models,
then the resulting estimator is consistent and asymptotically normal,
and standard techniques such as the bootstrap may be used to obtain
confidence intervals with correct coverage and hypothesis tests with
correct type 1 error control. Two issues arise with this approach.
First, with continuous or multi-valued exposures, it is often the case
that the research question requires estimating the effects
$\theta(\bar a, v)$ at multiple, possibly infinite, values $\bar a$.
Second, it is practically impossible to correctly pre-specify a
parametric model for equation (\ref{eq:defQ}), which means that this
estimation strategy will likely result in biased estimators.

In the case of a single time point ($\tau=1$), several methods exist
for tackling these problems
\citep[e.g.,][]{diaz2013targeted,kennedy2017non,westling2020unified,westling2020causal,colangelo2020double,
  semenova2021debiased,bonvini2022fast}. However, although some of
these methods deliver estimators with known asymptotic distributions,
none of them delivers root-n consistent inference in the case of
continuous exposures. Intuitively, the reason is that with continuous
exposures the parameter $\theta$ is not smooth as a functional of $\P$
in the sense that it is not pathwise differentiable \citep{Bickel97},
and thus root-n consistent estimation is not possible. More
importantly, these methods have not been generalized to the case of
multiple time points, and it remains unclear whether such
generalizations are possible.

In this article we adopt an alternative approach. Instead of targeting
$\theta(\bar a, v)$, we target its projection onto a parametric
working model, defined as follows.

\begin{definition}[Working marginal structural
  model]\label{def:wmsm} Let $\phi(\bar a, v)\in\Re^d$ denote a user-given
  transformation of $(\bar a, v)$, and let
  $m(\gamma \cdot \phi(\bar a, v))$ denote a parametric model for
  $\theta(\bar a,v)$ with parameter $\gamma$. We define the parameter
  of interest as the projection of $\theta(\bar a,v)$ onto the model,
  namely:
  \begin{equation}\label{eq:defbeta}
      \beta=\argmin_{\gamma\in\Re^d}\int L[\theta(\bar a, v), m(\gamma \cdot \phi(\bar a, v))]\dd\Lambda(\bar a,v),
  \end{equation} where $\Lambda(\bar a, v)$ is a user-given
  distribution with density function $\lambda(\bar a, v)$ and $L$ is a
  loss function that satisfies Assumption \ref{ass:canonical} below.
\end{definition}

The choice of the pair $(L,m)$ will be important for some of the
developments in this paper. In what follows we assume $(L,m)$
satisfies the following:
\begin{assumption}\label{ass:canonical}
  Assume $L$ and $m$ are such that
  \begin{equation*}
    \frac{\partial}{\partial \gamma}L[\theta(\bar a, v), m(\gamma
    \cdot \phi(\bar a, v))] = \{\theta(\bar a, v) - m(\gamma \cdot \phi(\bar a,
    v))\} \phi(\bar a, v).
  \end{equation*}
\end{assumption}
For example, for $m$ a logistic model we choose the cross-entropy loss
function $L(\theta, m)=-\theta\log m - (1-\theta)\log(1-m)$, for
linear $m$ we choose the quadratic loss function
$L(\theta,m) = (\theta-m)^2$, for log-linear $m$ we choose the
so-called Poisson loss-function $L(\theta, m)=-\theta\log m + m$,
etc. In general, for generalized linear models with canonical link,
the loss function derived from the negative log-likelihood loss is
guaranteed to satisfy Assumption \ref{ass:canonical}. Under regularity
conditions that allow exchanging the integral and the derivative,
$\beta$ is the solution to the estimating equation $\U(\gamma)=0$,
where
\begin{equation}
  \U(\gamma)=\int \{\theta(\bar a, v) - m(\gamma \cdot \phi(\bar a,
  v))\} \phi(\bar a, v)\dd\Lambda(\bar a, v).\label{eq:defU}
\end{equation}

We now discuss two simple estimation procedures based on inverse
probability weighting and regression adjustment. The efficient,
asymptotically normal estimators of \sec\ref{sec:eff} will rely on
insights from these estimators.

\section{Inverse probability weighting and g-computation
  (a.k.a. regression adjustment)}\label{sec:simple}

Putting together the definition of $\theta(\bar a, v)$ in equation
(\ref{eq:defQ}) with equation (\ref{eq:defU}) gives rise to the
following expression for the estimating equation:
\begin{equation}
  \U(\gamma)=\E\left[\left(\prod_{t=1}^\tau\r(A_t,H_t)\right)\{Y -
    m(\gamma \cdot \phi(\bar A, V))\} \phi(\bar A, V)\right],\label{eq:defIPW}
\end{equation}
which motivates the inverse probability weighting estimators of
\cite{Robinsetal00} that have become ubiquitous in applied research
using marginal structural models. Here $\r(a_t, h_t)$ is the density
ratio (often referred to as ``stabilized weights'') defined as
\[\r_t(a_t, h_t)=\frac{\lambda_t(a_t\mid \bar a_{t-1}, v)}{g_t(a_t\mid h_t)}.\]
Inverse probability weighting proceeds by obtaining an estimate
$\hat r$, and solving the estimating equation
\[\frac{1}{n}\sum_{i=1}^n\left(\prod_{t=1}^\tau\hat\r_t(A_{t,i},H_{t,i})\right)\{Y_i -
  m(\gamma \cdot \phi(\bar A_i, V_i))\} \phi(\bar A_i, V_i)=0\] in
$\lambda$, to obtain an estimate $\hat\beta_{\text{\scriptsize
    ipw}}$. Solutions to this estimating equation can be obtained
using standard methods for generalized estimating equations using
weights. If the marginal structural model is correct, the choice of
$\lambda$ affects the variance but not the consistency of the
estimators \citep{robins2000marginal}, but if the marginal structural
model is not correct and used merely as an approximation in the sense
of Definition \ref{def:wmsm}, then the choice of $\lambda$ changes the
projection and therefore the target estimand. When the weights are
estimated in a correctly specified parametric model, standard
Wald-type software output that ignores variability in estimation of
$r_t$ yields confidence intervals with conservative coverage
\citep[see e.g.,][Theorem 2.3]{vanderLaan2003}, and exact coverage may
be obtained with some additional calculations or using the
bootstrap. Because of the need for pre-specification of the model for
the weights, typical analyses with multivalued exposures assume $g$ and
$\lambda$ are distributions in simple exponential families such as
normal models with linear mean and constant variance \citep[e.g.,
][]{Robinsetal00}. This leads to misspecification in most applications
(e.g., skewed, heavy tailed, heteroscedastic exposures, or if the
dimension of $H_t$ is large), and therefore introduce bias into the
estimation procedure. Data-adaptive model selection techniques or
flexible advanced machine learning for conditional density estimation
\cite[e.g.][]{Diaz11, izbicki2017, dalmasso2020conditional} may be
employed to address this problem, but the theoretical foundation for
establishing general conservativeness/correctness of Wald-type and
bootstrap confidence intervals breaks down under data-adaptive
estimation of the weights. Developing estimators that can leverage
machine learning for estimation of the nuisance parameters (such as
the weights $r_t$) to alleviate model misspecification while retaining
approximately correct frequentist operating characteristics (e.g.,
coverage, type I error) motivates the developments of the next
section.

An alternative estimation strategy may be devised by expressing the
estimating equation $\U(\gamma)=0$ in terms of sequential regression
functions. To do so, we let $\U(\gamma)=\U_1+\U_2(\gamma)$, where
\begin{align*}
  \U_1 &= \int\theta(\bar a, v) \phi(\bar a, v)\dd\Lambda(\bar a, v),\\
  \U_2(\gamma) &= -\int m(\gamma \cdot \phi(\bar a, v)) \phi(\bar a, v)\dd\Lambda(\bar a, v),
\end{align*}
and define $\dot \U(\gamma) = \dot \U_2(\gamma)$ as the Jacobian of
$\U$ with respect to $\gamma$. Note that $\U_2(\gamma)$ does not
depend on $\P$, and that only $\U_1$, which does not depend on
$\gamma$, needs to be estimated. This property, which is a consequence of Assumption \ref{ass:canonical}, simplifies all estimation
procedures considerably. Specifically, it is only necessary to
construct an estimate $\hat \U_1$ and then obtain an estimate of
$\beta$ as the solution in $\gamma$ of $\hat \U_1+\U_2(\gamma)=0$. The
following lemma provides a sequential regression representation for
$\U_1$ that will be useful for this purpose:
\begin{lemma}[Sequential regression representation of $\U_1$]\label{lemma:seq}
  Initialize $\bar \T_{\tau+1}=Y\times \phi(\bar A, V)$. For $t=\tau,\ldots,1$,
  recursively define
  \begin{align}
    \T_t:(a_t, h_t)&\mapsto  \E[\bar \T_{t+1}(H_{t+1})\mid
                     A_t=a_t, H_t = h_t]\label{eq:seqreg}\\
    \bar \T_t:h_t&\mapsto \int \T_t(a_t, h_t)\dd\Lambda_t(a_t\mid \bar a_{t-1}, v)\notag
  \end{align}
  Then we have  $\U_1 = \E[\bar \T_1 (H_1)]$.
\end{lemma}

This is a direct consequence of (\ref{eq:defQ}) and the definition of
$\U_1$. This lemma motivates the construction of an alternative to
inverse probability weighting for marginal structural models: a
g-computation estimator using Newton-Raphson for root finding. This
estimator can be implemented in the following steps:
\begin{enumerate}[label=(\roman*)]
\item Compute an estimate $\hat \U_1$ by recursively (starting with
  $t=\tau$ and ending with $t=1$) fitting regressions for $\T_t$ and
  computing $\bar \T_t$ using numerical integration;
\item Initialize $k=0$ and $\hat \beta_k = \hat\beta_{\text{\scriptsize ipw}}$;
\item Let $\hat \beta_{k+1} = \hat\beta_k - [\dot
  \U_2(\hat\beta_k)]^{-1}[\U_2(\hat\beta_k)+\hat \U_1]$;\label{stepU11}
\item Update $k=k+1$\label{stepU12};
\item Iterate \ref{stepU11} and \ref{stepU12} until convergence.
\end{enumerate} This estimator, however, would suffer from similar
issues to the inverse probability weighted estimator. In particular,
if the regressions in (\ref{eq:seqreg}) are fitted using data-adaptive
regression estimators (e.g., machine learning), there is no general
theoretical basis to study the sampling distribution of the resulting
estimator of $\beta$, which leaves us without a theoretical foundation
upon which approximately correct confidence intervals and hypothesis
tests can be constructed. Nonetheless, the insights of this estimation
algorithm will be important for the development of the asymptotically
normal estimators of the next section.

\section{Efficiency theory}\label{sec:eff}

We now turn our attention to a discussion of efficiency theory for
estimation of $\beta$ in the nonparametric model. The efficient
influence function is an essential object that characterizes the
asymptotic behavior of all regular and efficient estimators
\citep[see the convolution theorem, e.g., in][]{Bickel97}; specifically: (i) the efficient influence function
can be used to construct locally efficient estimators; (ii) such
estimators often enjoys desirable properties such as double
robustness, which allows for some nuisance parameters to be
inconsistently estimated while preserving consistency of the estimator
of $\beta$; and (iii) asymptotic analysis of estimators constructed
using the efficient influence function often yields second-order bias
terms, which require slow convergence rates (e.g., $n^{-1/4}$) for the
nuisance parameters involved, thereby enabling the use of flexible
regression techniques in estimating these quantities. While efficient
estimators of $\beta$ based on the efficient influence function and
using flexible regression have been developed
\citep[e.g.,][]{rosenblum2010targeted, petersen2014targeted,
  schnitzer2014modeling}, they are only available for discrete
exposures and do not scale well to exposures taking on many values.

The efficient influence function is intimately related to a
first-order expansion of the parameter $\beta$ as a functional of the
data distribution $\P$, a so-called von-Mises expansion
\citep{mises1947asymptotic}. In what follows we present a study of the
efficient influence function and the von-Mises expansion which will
allow us to develop an estimator for continuous exposures that
leverages data-adaptive flexible regression for estimation of the
nuisance parameters and is root-n consistent and asymptotically normal
under slow-rate consistency of all the nuisance estimators.
Let $\eta=(\r_1,\T_1,\ldots,\r_\tau,\T_\tau)$ denote the vector of
nuisance parameters. In the following we will use a vector
$\eta'=(\r_1',\T_1',\ldots,\r_\tau',\T_\tau')$ that will typically
represent the probability limit of a given estimator $\hat \eta$. For
$t\leq \tau$, define the data transformation
\[ \D_t:z\mapsto \sum_{s=t}^\tau\left(\prod_{k=t}^s\r_k'(a_k,
    h_k)\right)\left\{\bar \T_{s+1}'(h_{s+1}) - \T_s'(a_s, h_s)\right\}
  + \bar \T_t'(h_t),\] where we will sometimes use
$\D_t(z;\eta'_t)$ to explicitly denote the dependence on
$\eta_t'=(\r_t',\T_t',\ldots,\r_\tau',\T_\tau')$. We have the
following result.
\begin{theorem}[von-Mises-type first order approximation]\label{lemma:von}
  Let $\D_{\tau+1}=Y\times \phi(\bar A, V)$, and define
  $\C_{t,s}'=\prod_{k=t+1}^{s-1}\r_k'(A_k,H_k)$, as well as {\small
    \[\rem_t(a_t, h_t, \eta_t') =
      \sum_{s=t+1}^\tau\E\left[\C_{t,s}'\{\r_s'(A_s,H_s)-\r_s(A_s,H_s)\}\{\T_s'(A_s,H_s)-\T_s(A_s,H_s)\}\,\,\bigg|\,\,
        A_t=a_t, H_t=h_t\right].\]} 
  We have
  \[\E[\D_{t+1}(Z;\eta_{t+1}')\mid A_t=a_t, H_t=h_t]=\T_t(a_t,h_t) +
    \rem_t(a_t,h_t;\eta'_t).\]
\end{theorem}
This theorem is analogous to Lemma 1 in \cite{luedtke2017sequential}
and Lemma 2 in \cite{rotnitzky2017multiply} for the standard g-formula
for dynamic regimes, and to Lemma 1 of \cite{diaz2021nonparametricmtp}
for longitudinal modified treatment policies. It shares important
connections to the von-Mises-type expansions used in some of the
semiparametric inference literature
\citep[e.g.,][]{mises1947asymptotic, vanderVaart98,
  robins2009quadratic}, and it has important implications which form
the basis of our estimation proposal. Specifically, note that if
$\eta_t'$ is such that, at each time point $s> t$ we have $\T_s'=\T_s$
or $\r_s'=\r_s$, then $\rem_t(a_t,h_t;\eta'_t)=0$. This implies that, for an
estimate ${\hat\eta}_{t+1}$, regressing $\D_{t+1}(Z;{\hat\eta}_{t+1})$ on $(A_t, H_t)$
provides a sequentially doubly robust estimator of $\T_t(a_t,h_t)$ in
the sense that it will be consistent if, for each time point $s>t$,
either $\T_s$ is estimated consistently or $\r_s$ is estimated
consistently.

Furthermore, the transformation $\D_t$ characterizes the efficiency
bound for estimation of $\beta$ in the following sense:
\begin{theorem}[Efficiency bound]\label{theo:eif}
  The random variable
  $\S(Z;\eta)= -[\dot \U_2(\beta)]^{-1}[\D_1(Z;\eta)-\U_1]$ is the
  efficient influence function for $\beta$ in the non-parametric
  model. Therefore, $\var[\S(Z;\eta)] $ is the local asymptotic minimax
  efficiency bound for estimation of $\beta$ in the sense that, for
  any estimator sequence $\beta_n$:
  \[\inf_{\delta
      >0}\liminf_{n\to\infty}\sup_{\Q:V(\Q-\P)<\delta}n\mathbb{E}\{\beta_n
    - \beta(\Q)\}^2\geq \diag\{\var_\P[\S_\P(Z;\eta_P)]\},\] where
  $V(\cdot)$ is the variation norm, $\mathbb{E}$ denotes expectation,
  and $\geq$ denotes element-wise inequality. We added indices $\P$
  and $\Q$ to emphasize sampling under $\P$ or $\Q$, and used notation
  $\beta(\Q)$ to denote the parameter computed at an arbitrary
  distribution.
\end{theorem}
This efficiency bound implies that, under sampling from distributions
$\Q$ in a shrinking neighborhood of the true probability distribution
$\P$, the worst-case asymptotic mean squared error of any estimator
sequence $\beta_n$ scaled by $n$ cannot be smaller than the variance
of the efficient influence function. Our goal is therefore to develop
estimators that achieve this bound.

\section{Construction of an efficient, sequentially doubly robust, and
  asymptotically normal estimator}

For $t=0$, inspection of Theorem~\ref{lemma:von} teaches us that it is
possible to construct an estimator of $\U_1$ by first computing an
estimator $\hat\eta$, and then averaging $\D_1(Z_i;\hat\eta)$ across
the sample. An estimator of $\beta$ can be obtained by plugging in
this estimator of $\U_1$ into the procedure for solving $\U(\gamma)=0$
detailed in \sec\ref{sec:simple}. If $\hat\eta$ is an estimator such
that $R_0(\hat\eta)=o_\P(1)$, Theorem~\ref{lemma:von} with $t=0$
implies that the resulting estimator is consistent. The consistency
condition $R_0(\hat\eta)=o_\P(1)$ can be achieved under the condition
that for each $t$, either $r_t$ or $T_t$ can be estimated
consistently, which in principle implies that the estimator is
consistent under $3^\tau$ out of $4^\tau$ configurations of
consistent/inconsistent estimation of each of the nuisance parameters
in $\eta$. Note, however, that the estimator of $\T_t$ in
\sec\ref{sec:simple} based on Lemma~\ref{lemma:seq} can only be
expected to be consistent in general when all of the estimators
$\T_s:s>t$ are also consistent. This implies that an estimator
na\"ively constructed by simply averaging $\D_1(Z_i;\hat\eta)$ will
only be consistent in $\alpha_\tau$ out of $4^\tau$ configurations of
consistent/inconsistent estimation of each of the nuisance parameters
in $\eta$, where $\alpha_\tau$ is the sequence defined by
$\alpha_k=6+2(\alpha_{k-1}-2)$ and $\alpha_1=3$. Note that $\alpha_\tau < 3^\tau$ for
$\tau>1$, so that the estimator would not leverage all the robustness
properties offered by Theorem~\ref{lemma:von}. In order to address
this problem, we will construct estimators of $\T_t$ based on
regressing the pseudo-outcome $\D_{t+1}(Z;\hat\eta_{t+1})$. Our
results below will guarantee that an estimator constructed in this way
is consistent under $3^\tau$ out of $4^\tau$ cases of
consistent/inconsistent estimation of the nuisance parameters.

Furthermore, proving asymptotic normality of an estimator of $\beta$
constructed as above would typically require that the nuisance
parameters $\eta$ are estimated within function classes of bounded
entropy, so that they satisfy functional versions of the central limit
theorem known as Donsker theorems \citep[see e.g., \sec2.5
of][]{vanderVaart&Wellner96}. These entropy conditions may limit the
kinds of estimators used in practice and impact the ability to use the
most flexible estimators to achieve the desired consistency of
$\hat\eta$. In order to avoid imposing entropy conditions, we use
sample splitting and cross-fitting
\citep{klaassen1987consistent,zheng2011cross,
  chernozhukov2018double}. Let ${\cal V}_1, \ldots, {\cal V}_J$ denote
a random partition of the index set $\{1, \ldots, n\}$ into $J$
prediction sets of approximately the same size. That is,
${\cal V}_j\subset \{1, \ldots, n\}$;
$\bigcup_{j=1}^J {\cal V}_j = \{1, \ldots, n\}$; and
${\cal V}_j\cap {\cal V}_{j'} = \emptyset$. In addition, for each $j$,
the associated training sample is given by
${\cal T}_j = \{1, \ldots, n\} \setminus {\cal V}_j$. We let
$\hat \eta_{j}$ denote the estimator of $\eta$ obtained by training
the corresponding prediction algorithm using only data in the sample
${\cal T}_j$. Further, we let $j(i)$ denote the index of the
validation set which contains observation $i$.

Having discussed all the building blocks, we are now ready to present
our proposed estimator. For any preliminary cross-fitted estimates
$\hat r_{1,j(i)},\ldots,\hat r_{\tau, j(i)}$, the estimator is defined
as follows:
\begin{enumerate}[label=Step \arabic*, align=left, leftmargin=*]
\item Initialize
  $\D_{\tau+1}(Z_i; \check\eta_{\tau,j(i)})= Y_i\times \phi(\bar A,V)$
  for $i=1,\ldots,n$.
\item For $t=\tau,\ldots,1$:
  \begin{enumerate}[label=(\roman*)]
  \item Compute the pseudo-outcome
    $\check Y_{t+1,i} =
    \D_{t+1}(Z_i;\check\eta_{t,j(i)})$ for all
    $i=1,\ldots,n$.
  \item For $j=1,\ldots,J$:
    \begin{itemize}
    \item Regress $\check Y_{t+1,i}$ on $(A_{t,i}, H_{t,i})$ using any
      regression technique and using only data points
      $i\in \mathcal T_{j}$.\label{step:2}
    \item Let $\check T_{t,j}$ denote the output, update
      $\check\eta_{t, j} = (\hat r_{t,j}, \check
      T_{t,j},\ldots,\hat r_{\tau,j}, \check T_{\tau,j})$, and
      iterate.
    \item Compute $\check{\bar T}_{t,j}$ by numerical integration or
      importance sampling.
    \end{itemize}
  \end{enumerate}
\item Define $\hat \U_1=n^{-1}\sum_{i=1}^n\D_1(Z_i;\check\eta_{j(i)})$.
\item Solve $\hat \U_1+\U_2(\gamma)=0$:
  \begin{enumerate}[label=(\roman*)]
  \item Initialize $k=0$ and $\hat \beta_k = \hat\beta_{\text{\scriptsize ipw}}$;
  \item Let $\hat \beta_{k+1} = \hat\beta_k - [\dot
    \U_2(\hat\beta_k)]^{-1}[\hat \U_1+\U_2(\hat\beta_k)]$;\label{step1}
  \item Update $k=k+1$\label{step2};
  \item Iterate \ref{step1} and \ref{step2} until convergence, i.e.,
    until $\U_2(\hat\beta_k)=\hat \U_1 + o_\P(n^{-1/2})$.
  \item Let
    $\hat\beta_{\text{\scriptsize sdr}}$ denote the resulting estimator.
  \end{enumerate}
\end{enumerate}

To prove the sequential double robustness and root-n consistency of
this estimator, it will be useful to have an alternative expression of
the second-order term $\rem_t$. Define the data-dependent
parameter
\[\check\T^0(a_t, h_t) = \E\big[\D_{t+1}(Z;\check\eta_t)\mid
  A_t=a_t,H_t=h_t\big],\] where the outer expectation is with respect
of the distribution $\P$, taking $\check\eta$
fixed. Theorem \ref{lemma:von} yields
\begin{equation}
  \T_t(a_t,h_t) = \check\T_t^0(a_t,h_t) +
  \rem_t(a_t, h_t;\check\eta).\label{eq:check}
\end{equation}
An induction argument yields the lemma below.
\begin{lemma}\label{lemma:remorder}
  Assume that $\P\{\r_t(A_t,H_t) < c\} =\P\{\hat\r_t(A_t,H_t) < c\}=1$
  for some $c<\infty$, and let $\check\T_t$ be the estimator defined
  above. Then
  \begin{equation}
    \rem_0(\check\eta) = \sum_{t=1}^\tau O_\P\big(||\hat\r_t -
    \r_t||\,||\check\T_t - \check\T_t^0||\big).\label{eq:rem2}
  \end{equation}
\end{lemma}
The proof of this lemma follows the same steps of the proof of Lemma 3
of \cite{diaz2021nonparametricmtp}. This representation of the
remainder term reveals that each component of $\rem_0(\check\eta)$ at
time $t$ depends solely on regressions fit at that time point, and
avoids the dependence of $T_t$ on $T_s:s>t$ implied by its sequential
regression definition. This representation is thus more useful to
establish sequential doubly robust consistency and asymptotic
normality. In particular, we have:
\begin{theorem}[Weak convergence of $\hat\beta_{\text{\scriptsize sdr}}$]\label{theo:astmle}
  Assume that
  $\sum_{t=1}^\tau||\hat\r_t - \r_t||\, ||\check\T_t - \check\T_t^0||
  = o_\P(n^{-1/2})$ and that
  $\P\{\r_t(A_t,H_t) < c\}=\P\{\hat\r_t(A_t,H_t) < c\}=1$ for some
  $c<\infty$. Then
  \[n^{1/2}(\hat\beta_{\text{\scriptsize sdr}} - \beta)
    \rightsquigarrow N(0,\Sigma),\] where
  $\Sigma=\var\{\S(Z;\eta)\}$ is the non-parametric efficiency
  bound. Therefore, $\hat\beta_{\text{\scriptsize sdr}}$ is efficient
  in the sense of Theorem~\ref{theo:eif}.
\end{theorem}
The above theorem shows that $\hat\beta_{\text{\scriptsize sdr}}$ is
efficient and provides the conditions under which we can compute
Wald-type correct confidence intervals and hypothesis tests.  The
following proposition shows that the estimator is sequentially doubly
robust in the sense that it is consistent if, at each time point,
either $\check\T_t$ is consistent for $\check\T_t^0$, or if
$\hat \r_t$ is consistent for $\r_t$.
\begin{prop}[Sequential doubly robust consistency of
  $\hat\beta_{\text{\scriptsize sdr}}$]\label{theo:conssdr} Assume
  that, for each time $t$, either $||\hat\r_t - \r_t|| = o_\P(1)$ or
  $||\check\T_t - \check\T^0_t|| = o_\P(1)$ . Then we have
  $\hat\beta_{\text{\scriptsize sdr}} =\beta + o_\P(1)$.
\end{prop}

The second condition of Theorem~\ref{theo:astmle} is standard in
causal inference, simply stating that there is enough experimentation
in the treatment mechanism such that $g(a_t\mid h_t)$ is positive
whenever $\lambda(a_t\mid \bar a_{t-1}, v)$ is positive. Note that,
since $\lambda$ is given by the user and forms part of the projection
in the definition of $\beta$, this assumption may be arranged by
definition if the areas of poor support of $A_t$ are known a-priori.
The assumption regarding consistency of the nuisance estimators can be
satisfied, for example, if all the nuisance parameters converge to
their true values at $n^{1/4}$-rate. The required rates are achievable
by many data-adaptive regression algorithms. See, for example,
\citet{bickel2009simultaneous} for rate results on $\ell_1$
regularization, \citet{wager2015adaptive} for rate results on
regression trees, \citet{zhang2005boosting} for boosting,
\citet{chen1999improved} for neural networks, and
\cite{benkeser2016highly} for the highly adaptive lasso. Stacking or
ensemble learners such as the Super Learner~\citep{vdl2007super},
which have additional model selection properties such as oracle
guarantees, may also be used.

\section{Numerical studies}

To study the sequential dual robustness and empirical performance of our proposed method, we conducted a simulation study with the following data-generating mechanism. We generate datasets with $\tau = 4$ time points, with treatment generated at each time point drawn from a multinomial distribution with five possible outcomes $\{0,\ldots,4\}$, denoted as $A_t \mid H_t \sim \text{Multinomial}(k = 5, p_t = f_A(H_t))$, where $H_t = (A_{t-1}, L_{t}, A_{t-2}, L_{t-1})$ varies with time and \[ f^*_{iA}(H_t) = \exp\left(c_{1it} + c_{2it} A_{t-1}/4 + c_{3it} L_t + c_{4it} A_{t-2}/4 + c_{5it}  A_{t-1}/4 L_t + c_{6it} L_{t-1} \right), \] 
and 
\[ f_{iA}(H_t) = f^*_{iA}(H_t) / \sum_{j = 0}^{4} f^*_{jA}(H_t), \]
with $\{c_{1it},c_{2it},c_{3it},c_{4it},c_{5it},c_{6it}\}$ constants at each time point for $i \in \{0,4\}$ corresponding to each possible values of $A_t$, leading to $f_A(H_t)$ being a vector of length 5. The initial covariate $L_1$ was generated from a univariate discrete distribution with five categories, while all subsequent time-varying covariates followed binomial distributions, $L_t \mid H_t \sim \text{Binomial}(p_t = f_L(H_t))$, where $H_t = (A_{t-1}, L_{t-1}, L_{t-2})$ and \[ f_L(H_t) = 1/ \left( 1 + \exp \left( - \left( -.5 \times A_{t-1} + L_{t-2} +  2 \times L_{t-1} - A_{t-1} \times L_{t-1} + L_{t-2} \times L_{t-1} \right) \right) \right). \] The outcome variable $Y$ was also binomially distributed, $Y \mid H_t \sim \text{Binomial}(p = f_Y(H_t))$, where $H_t = (A_4, L_4, A_3, L_3, A_2, L_2, A_1, L_1)$ and $f_Y(H_t) = 1/(1+\exp(-\alpha)) $ with \begin{align*}
    \alpha =& -2.5 + 5 \times (L_4 + L_3 + L_2 + L_1/5) + 5 \times L_4 \times L_3 \times L_2 \times L_1/5 - \\
    & A_4 \times (5 - 2 \times A_3 \times L_4 \times L_3) - A_3 \times (4 - 1.5 \times A_2 \times L_3 \times L_2) - \\
    & A_2 \times (3 - A_1 \times L_2 \times L_1/5) - A_1 \times (2 - L_1/5).
\end{align*}The MSM we selected is a simple model with an intercept and a slope for the cumulative treatment, as $\phi(\bar{a}, v) = (1, \sum_{t=1}^{\tau = 4} a_t)$. We define the working MSM $m(\beta \cdot \phi(\bar a , v) )$ where $\beta$ is the projection of $\theta(\bar{a},v)$ onto the model and is defined as Eq. (\ref{eq:defbeta}),
where $m(\beta \cdot \phi(\bar a , v) ) = 1/(1 + \exp(-\beta \cdot \phi(\bar a , v) ))$ is a logistic model and $L[\theta(\bar a, v), m(\gamma \cdot \phi(\bar a, v))] = -\theta\log m(\gamma \cdot \phi(\bar a , v) - (1-\theta(\bar a, v))\log(1-m(\gamma \cdot \phi(\bar a , v)) $ is the cross-entropy loss function, satisfying Assumption \ref{ass:canonical}. Note that here we constructed $\Lambda(\bar a, v)$ as a multinomial distribution to be as close as possible to the true marginal distribution of the treatment. We use a simple model with only an intercept and a term for cumulative treatment, i.e., we let the linear predictor be $\phi(\bar{a},v) = (1, \sum_{t=1}^4 a_t)$.

While the methodology presented here is a general framework that can accommodate any type of treatment distribution, including continuous ones, in practice this requires the estimation of conditional densities of the exposure in the weights calculation step. Currently, the statistical and machine learning literature are limited in the development of such estimators, with only a few options available. In contrast, the literature for data-adaptive estimators of probability mass functions, particularly classifiers, is well developed in comparison. Therefore, to illustrate the performance of our proposed algorithm and avoid possible issues in estimation of continuous densities, which are orthogonal to our proposal, we use a discrete exposure in this simulation. Notably, even in this simple setting, existing non-parametric estimators such as targeted maximum likelihood estimation (TMLE) \citep{rosenblum2010targeted}, its extended version for dynamic longitudinal MSMs \citep{petersen2014targeted} would not be applicable as they would require fitting very large models in datasets of size $5^4 \times n$, which would be computationally prohibitive for most regression algorithms even relatively in small sample sizes.

To approximate the true value of parameter $\beta$ in the MSM under this data-generating mechanism, we created a large dataset ($n = 10^6$) and applied the Inverse Probability Weighting (IPW) estimator with true weights, yielding a cumulative treatment odds of -0.21. We then simulated $S = 200$ datasets under this setup for various sample sizes (n = 250, 500, 1000, 2000, 3000, 4000, 5000), comparing our Sequential Doubly-Robust (SDR) estimator against a targeted maximum likelihood estimator-like (TMLE) estimator (details in the appendix \sec\ref{tmle-est}) and the IPW estimator, across different scenarios (see Table \ref{tab:tab1} for details). To explore the performance of these estimators under mis-specified models for both treatment probability mass and outcome, we used the mlr3superlearner and SuperLearner libraries in R. To obtain scenarios under inconsistent estimators of the nuisance parameters, we use a strong misspecification that ignores all covariates and uses marginal empirical averages as predictors. The scenarios assessed are detailed in Table \ref{tab:tab1}. The performance of each estimator was evaluated based on Monte Carlo approximation to the bias $(1/S \times \sum_{s=1}^S (\hat{\beta_s} - \beta))$, scaled bias $(1/S \times \sum_{s=1}^S (\hat{\beta_s} - \beta) \times \sqrt{n})$ , the 95\% coverage $(1/S \sum_{s=1}^S I( \beta \in (\hat{\beta_s}^{\text{low}} , \hat{\beta_s}^{\text{high}})))$ where $\hat{\beta_s}^{\text{low}}$ and $\hat{\beta_s}^{\text{high}}$ are respectively the lower and upper bounds of the 95\% confidence interval, and the mean squared error $(1/S  \times \sum_{s=1}^S (\hat{\beta_s} - \beta)^2 \times n)$. The results summarized in Figure \ref{fig:Figure1} and Table \ref{tab:tab3} (in the appendix \sec\ref{appsec:simres}).

\begin{table}[ht]

\caption{Details on the numerical study scenarios. For data-adaptive estimation of the outcome model we used a stack of learners with the R package SuperLearner (methods used: mean, glmnet, glm.interaction, earth) and are annotated as 'SL' in the table. For data-adaptive estimation of the probability mass function used for weigths estimation we relied on the R package mlr3superlearner (methods used: mean, lightgbm, multinom, xgboost, nnet, knn, rpart, naivebayes, glmnet, randomforest, ranger) and are annotated as 'mlr3' in the table. The mis-pecified case where modeled by making the stacks described above only contain the mean model, annotated as 'Mean' in the table. Of note the R libraries used for the outcome model and probability mass model differ because the superLearner package currently does not handle categorical outcomes.}
\centering
\begin{tabular}[t]{l|l|l|l|l|l|l|l|l}
\hline
\multicolumn{1}{c|}{ } & \multicolumn{4}{c|}{Outcome Model} & \multicolumn{4}{c}{Probability Mass Model} \\
\cline{2-5} \cline{6-9}
  & T=1 & T=2 & T=3 & T=4 & T=1 & T=2 & T=3 & T=4\\
\hline
Scenario 1 & SL & SL & SL & SL & mlr3 & mlr3 & mlr3 & mlr3\\
\hline
Scenario 2 & SL & SL & SL & SL & Mean & Mean & Mean & Mean\\
\hline
Scenario 3 & Mean & Mean & Mean & Mean & mlr3 & mlr3 & mlr3 & mlr3\\
\hline
Scenario 4 & SL & SL & Mean & Mean & Mean & Mean & mlr3 & mlr3\\
\hline
Scenario 5 & Mean & Mean & SL & SL & mlr3 & mlr3 & Mean & Mean\\
\hline
\end{tabular}
\label{tab:tab1}
\end{table}

\begin{figure}[!ht]
    \centering
    \includegraphics[width=1\textwidth]{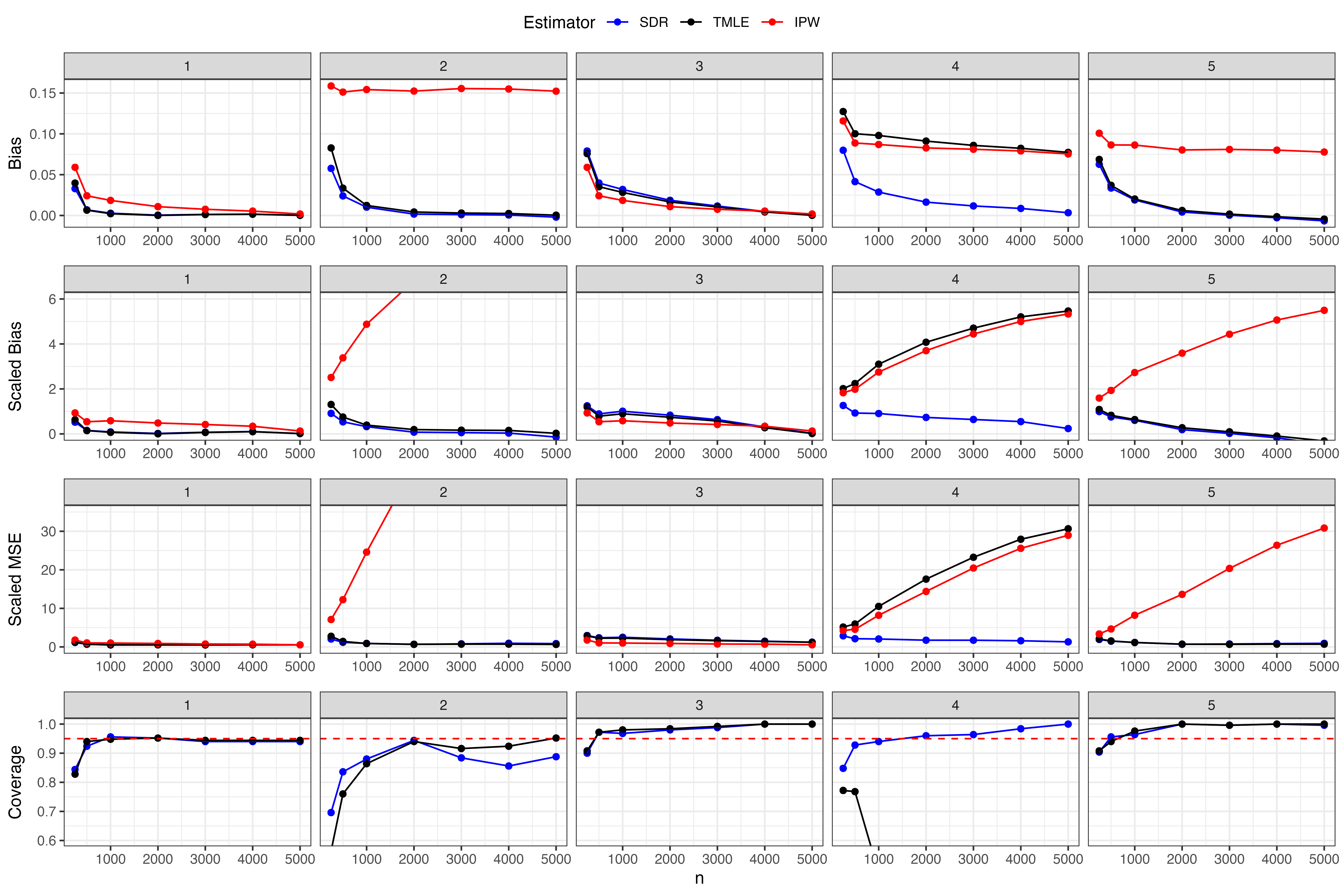}
    \caption{Summary of Simulation Results. The grey box with a number corresponds to the scenario detailed in Table \ref{tab:tab1}. The top row of plots illustrates the distribution of mean bias, the second row represents the scaled bias. The third row depicts the mean squared error scaled by the sample size. The bottom row shows the coverage probability of the true value within our 95\% confidence intervals, the dashed red bar representing the expected coverage of 0.95.}
    \label{fig:Figure1}
\end{figure}

Our simulation results demonstrate that under the specified data generating mechanism, our Sequentially Doubly Robust (SDR) estimator outperforms existing methods such as Inverse Probability Weighting (IPW) and the Targeted Maximum Likelihood Estimation-like (TMLE) in terms of our selected metrics. These simulations were structured to evaluate the estimator's robustness and consistency across a variety of scenarios involving distinct model mis-specifications for the weighting and outcome models, specifically designed to test double-robustness attributes. As expected, in Scenario 1, using flexible data-adaptive methods for nuisance parameter estimation our estimator exhibits a scaled bias approaching 0, a scaled MSE approaching the efficiency bound $\diag\{\var_\P[\S_\P(Z;\eta_P)]\}$ described in Theorem \ref{theo:eif}, and a coverage approaching 95\% with increasing sample size. Following Proposition \ref{theo:conssdr}, we observe a bias converging to 0 with increasing sample size for Scenarios 2-5. Thus numerically demonstrating the sequentially doubly-robust properties of our proposed estimator, unlike the TMLE and IPW estimators that do not show this behavior. The R code for the numerical study is available on \href{https://github.com/AxelitoMartin/lmsm_simulations}{GitHub}.

\section{Illustrative application}

To illustrate the proposed methodology, we used a contemporary and publicly accessible dataset on COVID-19. This dataset encompasses longitudinal data that provides a comprehensive summary of the impact of state-level lockdown and masking mandates on COVID-10 outcomes across counties in the USA. This a relevant dataset in the scope of our proposed methodology as various states implemented different policies to control the spread of the virus. The study spans the calendar period corresponding to the peak of the pandemic, from early 2020 to late 2021, a critical time period in fluctuations in the number of COVID-19 cases and actions taken by state governments. The dataset includes multiple measures of time-varying mobility indexes reflecting changes in population movement patterns derived from mobile device data, the state-level masking mandates, along with demographic data and healthcare infrastructure statistics that could affect the number of COVID-19 cases. Previous research by Wong \citep{wongcovid2022} used a marginal structural model to demonstrate that an increase in the mobility index is associated with a subsequent rise in the incidence of new COVID-19 cases two weeks later. This finding underpins our use of a sequential doubly-robust estimator to hypothesize that increased mobility exerts a positive causal effect on the cumulative incidence of COVID-19 cases across multiple time points. 

The dataset had the following structure: $L_t=f_{L_t}(A_{t-1}, H_{t-1})$, $A_t=f_{A_t}(H_t)$, and $Y=f_Y(A_\tau, H_\tau)$, where each value of $t$ represents a calendar week. $H_t$ is comprised of a set of baseline covariates that are static, containing demographic data and healthcare infrastrucure statistics, and a set of time-varying covariates containing the prior exposures values of mobility index, the evolving masking mandates and critically the prior values of the outcome of interest. We used the cumulative number of observed COVID-19 cases as our outcome, thus prior outcomes are also predictors of future outcomes and were included as time-varying confounders. For illustrative purposes, we constructed a dataset with eight time points ($\tau=8$), over the first 6 months of the pandemic. To ensure the dataset generated adhered to the temporal requirements essential for causal inference frameworks, we introduced a three-week interval between each mobility measurement considered, the final outcome is the cumulative number of COVID-19 cases observed two weeks after the last observed exposure value. This interval allows for the assessment of the impact of intervening public health interventions, such as changes in mobility restrictions and masking mandates. In our analysis, we categorized the mobility index into five discrete groups, each representing 20\% of the data range and labeled from 0 to 4. To mimic our simulation study we used a MSM with a simple model with an intercept and a slope for the average exposure, as $\phi(\bar{a}, v) = (1, \sum_{t=1}^{\tau = 8} a_t/\tau)$. Our findings, summarized in Table \ref{tab:tab2} encompassing different starting weeks, corroborate the hypothesized relationship where higher levels of mobility are significantly associated with an increase in the cumulative number of COVID-19 cases, aligning with conclusions drawn in prior studies. We report the value for our proposed estimator ($\hat{\beta}_{sdr}$), with its variance ($\hat{\sigma}_{\hat{\beta}_{sdr}}^2$) and 95\% confidence interval ($\hat{\beta}_{sdr}^{low}$, $\hat{\beta}_{sdr}^{high}$). For completion we also report the estimated values of the parameter when using the TMLE estimator ($\hat{\beta}_{TMLE}$), with its variance ($\hat{\sigma}_{\hat{\beta}_{TMLE}}^2$) and 95\% confidence interval ($\hat{\beta}_{TMLE}^{low}$, $\hat{\beta}_{TMLE}^{high}$) and the IPW estimator ($\hat{\beta}_{IPW}$). 

\begin{table}[!ht]
\centering
\caption{Results of the mean effect of mobility on the cumulative number of COVID-19 within the first 6 months of the pandemic.}
\begin{tabular}{l|rrrrrrrrr}
\toprule
& $\hat{\beta}_{sdr}$ & $\hat{\sigma}_{\hat{\beta}_{sdr}}^2$ & $\hat{\beta}_{sdr}^{low}$ & $\hat{\beta}_{sdr}^{high}$ & $\hat{\beta}_{TMLE}$ & $\hat{\sigma}_{\hat{\beta}_{TMLE}}^2$ & $\hat{\beta}_{TMLE}^{low}$ & $\hat{\beta}_{TMLE}^{high}$ & $\hat{\beta}_{IPW}$ \\
\midrule
Intercept & 2036.6 & 4527.2 & 1904.8 & 2168.6 & 2065.4 & 5541.5 & 1919.5 & 2211.3 & 2000.1 \\
Slope & 297.6 & 465.9 & 255.3 & 339.9 & 300.2 & 854.8 & 242.9 & 357.5 & 285.7 \\
\bottomrule
\end{tabular}
\label{tab:tab2}
\end{table}

\section{Discussion}

In this manuscript, we have used semi-parametric efficiency theory to propose a novel sequentially doubly-robust estimator and asymptotically normal. Moreover, these properties hold even when relying on flexible data-adaptive methods to estimate the relevant nuisance parameters. These theoretical properties were validated through both a detailed simulation study and a real-world case study, with results corroborating those of previous investigations. Our methodology addresses a significant gap in the literature related to complex longitudinal studies, which often grapple with a high dimensional space of potential pathways. This 'curse of dimensionality' typically renders many estimators unserviceable, yet our approach retains practical properties that facilitate robust inference. The potential applications of our method, particularly in the field of health research, are extensive and promising.

However, during the implementation of our proposed algorithm, we encountered several challenges, particularly from numerical and computational perspectives. The task of repeatedly estimating densities and guaranteeing their product remains numerically stable has proven to be complex. To address this, we intend to incorporate Riesz representers \citep{chernozhukov2022riesznet} in our future work, which we anticipate will help bypass this hurdle. The iterative integration required in the outcome model imposes a substantial computational burden. Current approximations, such as Riemann integration, may introduce biases due to inadequate coverage of the integration domains. Although this issue can be potentially resolved in a simulated environment by specifying a predefined integration region, a definitive solution for real-world data applications remains elusive. Finally, identifying the optimal solution $\gamma$ such that $U_2(\gamma) = U_1$ depends on the initial guess and the optimization method employed. This process can be intricate and sensitive to the chosen parameters and techniques.

\newpage
\bibliographystyle{plainnat}
\bibliography{main}
\newpage
\appendix

\input{sm}

\end{document}

%% file: _preamble.tex
\usepackage[inline,shortlabels]{enumitem}
\usepackage{float}
\usepackage{mathtools}
\usepackage{subdepth}
\usepackage{graphicx}
\usepackage[round]{natbib}
\usepackage[margin=1in]{geometry}
\usepackage{tikz}
\usetikzlibrary{arrows.meta}
\usetikzlibrary{decorations.markings}
\usepackage[linesnumbered]{algorithm2e}
\RestyleAlgo{ruled}

\tikzset{+ /.tip = {Bar[sep=-3pt 2,width=3pt 4]_[sep=0]}}
\usepackage[english]{babel}
\usepackage{longtable}
\usepackage{color}
\usepackage{mathtools}
\usepackage{amssymb,amsmath,amsthm}
\usepackage{multirow}
\usepackage[titletoc,title]{appendix}
\usepackage{authblk}
\usepackage{bbm}
\usepackage{setspace}
\usepackage{dsfont}
\usepackage[OT1]{fontenc}
\usepackage{subcaption}
\usepackage{refcount}
\usepackage{booktabs}
\graphicspath{ {../simulations/01_init_manuscript/graphs/} }

\usepackage{xr-hyper}
\usepackage[colorlinks,citecolor=blue,urlcolor=blue]{hyperref}

\newtheorem{prop}{Proposition}
\newtheorem{theorem}{Theorem}

\AtEndDocument{\refstepcounter{theorem}\label{finalthm}}
\AtEndDocument{\refstepcounter{equation}\label{finaleq}}

{
  \theoremstyle{definition}
  \newtheorem{assumption}{Assumption}
}

{
  \theoremstyle{definition}
  
}

\DeclareMathOperator{\diag}{diag}

\newtheorem{lemma}{Lemma}
\AtEndDocument{\refstepcounter{lemma}\label{finallemma}}

\newtheorem{definition}{Definition}

 \DeclareMathOperator{\var}{\mathsf{Var}}

\renewcommand{\P}{P}

\newcommand{\Q}{\mathsf{Q}}

\newcommand{\rem}{\mathsf{R}}
\newcommand{\g}{g}

\renewcommand{\r}{\mathsf{r}}

\let\sec\S 

\newcommand{\U}{\mathsf{U}}

\newcommand{\C}{\mathsf{K}}
\renewcommand{\H}{\mathsf{H}}

\newcommand{\indep}{\mbox{$\perp\!\!\!\perp$}} 
 \newcommand{\dd}{\,\mathrm{d}}
\newcommand{\Pn}{\mathsf{P}_n}

\newcommand{\Pnj}{\mathsf{P}_{n,j}}
\newcommand{\Gnj}{\mathsf{G}_{n,j}}
\newcommand{\Gn}{\mathsf{G}_{n}}

\newcommand{\D}{\mathsf{D}}
\renewcommand{\S}{\mathsf{S}}
\newcommand{\T}{\mathsf{T}}

\newcommand{\E}{\mathsf{E}}
\renewcommand{\P}{\mathsf{P}}

\usepackage{accents}

\renewenvironment{proof}{{\it Proof }}{\qed \\}
\DeclareMathOperator*{\argmin}{\arg\!\min}
\DeclarePairedDelimiterX{\norm}[1]{\lVert}{\rVert}{#1}

\pgfdeclarelayer{background}
\pgfsetlayers{background,main}
\usetikzlibrary{arrows,positioning}
\tikzset{
>=stealth',
punkt/.style={
rectangle,
rounded corners,
draw=black, very thick,
text width=6.5em,
minimum height=2em,
text centered},
pil/.style={
->,
thick,
shorten <=2pt,
shorten >=2pt,}
}
\newcommand{\Vertex}[3]
{\node[minimum width=0.6cm,inner sep=0.05cm] (#2) at (#1) {#3};
}
\newcommand{\Vertexr}[3]
{\node[rectangle, draw, minimum width=0.6cm,inner sep=0.05cm] (#2) at (#1) {#2};
}
\newcommand{\ArrowR}[3]%
{ \begin{pgfonlayer}{background}
\draw[->,#3] (#1) to[bend right=30] (#2);
\end{pgfonlayer}
}
\newcommand{\ArrowLW}[3]%
{ \begin{pgfonlayer}{background}
\draw[->,#3] (#1) to[bend left=30] (#2);
\end{pgfonlayer}
}

\newcommand{\ArrowL}[3]%
{ \begin{pgfonlayer}{background}
    \draw[->,#3] (#1) to[bend left=45] (#2);
  \end{pgfonlayer}
}

\newcommand{\EdgeL}[3]%
{ \begin{pgfonlayer}{background}
\draw[dashed,#3] (#1) to[bend right=-45] (#2);
\end{pgfonlayer}
}

\newcommand{\Arrow}[3]%
{ \begin{pgfonlayer}{background}
\draw[->,#3] (#1) -- +(#2);
\end{pgfonlayer}
}

\allowdisplaybreaks
\pgfarrowsdeclare{arcs}{arcs}{...}
{
  \pgfsetdash{}{0pt} 
  \pgfsetroundjoin   
  \pgfsetroundcap    
  \pgfpathmoveto{\pgfpoint{-5pt}{5pt}}
  \pgfpatharc{180}{270}{5pt}
  \pgfpatharc{90}{180}{5pt}
  \pgfusepathqstroke
}
\newcommand{\ArrowB}[3]%
{ \begin{pgfonlayer}{background}
    \draw[|-arcs,line width=0.4mm,shorten <= 0.3cm,shorten >= 0.3cm,#3] (#1) -- +(#2);
  \end{pgfonlayer}
}

\newcommand{\titlepaper}{Non-parametric efficient estimation of
  marginal structural models with multi-valued time-varying treatments}

\date{\today}
\author[1]{Axel Martin\thanks{Corresponding author: Axel.Martin@nyulangone.org}}
\author[1]{Michele Santacatterina}
\author[1]{Iv\'an D\'iaz}

\affil[1]{\small Division of Biostatistics, Department of Population Health, New York University Grossman School of Medicine.}

%% file: sm.tex
\setcounter{equation}{\getrefnumber{finaleq}}
\addtocounter{equation}{-2}
\setcounter{theorem}{\getrefnumber{finalthm}}
\addtocounter{theorem}{-2}

\appendix
\section*{Supplementary Material}
\addcontentsline{toc}{section}{Supplementary Material}

\section{von-Mises-type first order approximation (Lemma \ref{lemma:von})}
\begin{proof}
  This lemma follows from recursive application for
  $s=t+1,\ldots,\tau$ of the following
  relationship:
  \begin{align*}
    \textcolor{blue}{\E[\bar T_{s,1}}&\textcolor{blue}{(H_s) - \bar T_{s}(H_s)\mid A_{s-1},
                                       H_{s-1}]} =\\
    = -&\E\left\{\frac{\lambda_{s}(A_s\mid
         \bar A_{s-1}, V)}{\g_{s,1}(A_s\mid H_s)}[\bar T_{s+1, 1}(
         H_{s+1}) - T_{s, 1}(A_s,
         H_s)]\,\,\bigg|\,\,A_{s-1},H_{s-1}\right\}\\
    +&\E\left\{\left[\frac{\lambda_s(A_s\mid
       \bar A_{s-1}, V)}{\g_s(A_s\mid H_s)}-\frac{\lambda_{s}(A_s\mid
       \bar A_{s-1}, V)}{\g_{s,1}(A_s\mid H_s)}\right][T_{s,1}(A_s,
       H_s) - T_s(A_s,
       H_s)]\,\,\bigg|\,\,A_{s-1},H_{s-1}\right\}\\
    +&\E\left\{\frac{\lambda_s(A_s\mid
       \bar A_{s-1}, V)}{\g_{s,1}(A_s\mid
       H_s)}\textcolor{blue}{\E[\bar T_{s+1, 1}(
       H_{s+1}) - \bar T_{s+1}(
       H_{s+1})\mid A_s, H_s]}\,\,\bigg|\,\,A_{s-1},H_{s-1}\right\},
  \end{align*}
  which follows because
\begin{align*}
    \E&\left[\frac{\lambda_s(A_s\mid
                    \bar A_{s-1}, V)}{\g_s(A_s\mid H_s)}T_{s,1}(A_s,H_s)\,\,\bigg|\,\, A_{s-1},
                    H_{s-1}\right]\\
                  &=\E\left[\E\left\{\frac{\lambda_s(A_s\mid
                    \bar A_{s-1}, V)}{\g_s(A_s\mid
                    H_s)}T_{s,1}(A_s,H_s)\,\,\bigg|\,\, H_s\right\}\,\,\bigg|\,\, A_{s-1},
                    H_{s-1}\right]\\
                  &=\E[\bar T_{s,1}(H_s)\mid A_{s-1},
                    H_{s-1}].
  \end{align*}
\end{proof}

\section{Efficient influence function (Theorem \ref{theo:eif})}

\begin{proof}
  In this proof we will use $\beta(\P)$ and $\theta(\P_\epsilon)$ to
  denote the parameters of interest as functionals that map the
  distribution $\P$ in the model to real numbers. The function
  $\S(Z;\eta)$ is the EIF of $\beta(\P)$ if it satisfies
  \begin{equation}
    \frac{\dd}{\dd  \epsilon}\beta(\P_\epsilon)\bigg|_{\epsilon=0} =
    \E[\S(Z;\eta) h(Z)],\label{eq:defeif}
  \end{equation}
  where $\P_\epsilon$ is a smooth parametric submodel
  with $\P_{\epsilon=0}=\P$ that locally
  covers the non-parametric model, with score
  \[h(Z)=\left(\frac{\dd \log \P_\epsilon}{\dd
        \epsilon}\right)\bigg|_{\epsilon = 0}.\]
  Define \[\H(\epsilon, \gamma) = \int\{\theta(\P_\epsilon)(\bar a, v)
    - m(\gamma\cdot \phi(\bar a, v))\}\phi(\bar a,
    v)\dd\Lambda(\bar a, v),\]
  and let \[\dot \H_\gamma = \frac{\dd
      \H}{\dd\gamma};\quad \dot \H_\epsilon = \frac{\dd
      \H}{\dd\epsilon},\]
  where we note that $\dot \H_\gamma(0,\beta) = \dot \U_2(\beta)$. The
  implicit function theorem applied to $\H(\epsilon,
  \beta(\P_\epsilon))=0$ shows that
  \[\frac{\dd}{\dd  \epsilon}\beta(\P_\epsilon)\bigg|_{\epsilon=0}
    =-[\dot \U_2(\beta(P))]^{-1}\dot \H_\epsilon(0,\beta(\P)).\]
  It remains to show that $\dot \H_\epsilon(0,\beta(P)) =
  \E\{[\D_1(Z;\eta)-\U_1]h(Z)\}$. To prove this, notice that
  \[\dot \H_\epsilon(0,\gamma) =\frac{\dd}{\dd\epsilon}\int
    \theta(\P_\epsilon)(\bar a, v)\phi(\bar a,
    v)\dd\Lambda(\bar a, v)\,\bigg|_{\epsilon=0}=\frac{\dd}{\dd\epsilon}U_1(\P_\epsilon)\,\bigg|_{\epsilon=0}.\]
  Theorem~\ref{lemma:von} implies that
  \[ \U_1(\P_\epsilon) = \U_1(\P) + \int \{\D_1(Z;\eta) - \U_1(\P)\} \dd\P_\epsilon - \rem_0(\eta_\epsilon,\eta),\]
  where we denote
  \[ \rem_0(\eta',\eta) = \sum_{s=1}^\tau\E\left[\{\r_s'(A_s, H_s) -
      \r_s(A_s, H_s)\}\{\T_s'(A_s, H_s) - \T_s(A_s, H_s)\} \right].\]
  Differentiating with respect to $\epsilon$ and evaluating at
  $\epsilon=0$ yields
  \begin{align*}
    \frac{\dd }{\dd \epsilon}\U_1(\P_\epsilon)\bigg|_{\epsilon = 0}
    &= \int\{\D_1(Z;\eta) - \U_1(\P)\}
      \left(\frac{\dd  \P_\epsilon}{\dd \epsilon}\right)\bigg|_{\epsilon
      = 0} - \frac{\dd }{\dd
      \epsilon}\rem_0(\eta_\epsilon,\eta)\bigg|_{\epsilon = 0} \\
    &= \int\{\D_1(Z;\eta) - \U_1(\P)\}
      \left(\frac{\dd \log \P_\epsilon}{\dd
      \epsilon}\right)\bigg|_{\epsilon = 0}\dd \P - \frac{\dd }{\dd \epsilon}\rem_0(\eta_\epsilon,\eta)\bigg|_{\epsilon = 0},
  \end{align*}
  and the expression for the efficient influence function follows after noticing that
  \[\frac{\dd }{\dd
      \epsilon}\rem_0(\eta_\epsilon,\eta)\bigg|_{\epsilon = 0}=0.\]
  The second part of the theorem regarding the efficiency bound
  follows from Corollary 2.6 of \cite{vandervaart2002}.
\end{proof}

\section{Asymptotic Normality of TMLE (Theorem~\ref{theo:astmle})}
\begin{proof}
  We will first proof that $\hat \U_1$ is an asymptotically linear
  estimator of $\U_1$. We will then use standard M-estimation theory
  to prove the result of the theorem.
  
  Let $\Pnj$ denote the empirical distribution of the prediction set
  ${\cal V}_j$, and let $\Gnj$ denote the associated empirical process
  $\sqrt{n/J}(\Pnj-\P)$. Let $\Gn$ denote the empirical process
  $\sqrt{n}(\Pn-\P)$.  We use $E(g(Z_1,\ldots,Z_n))$ to denote
  expectation with respect to the joint distribution of
  $(Z_1,\ldots,Z_n)$ (as opposed to the script letter $\E$ used to
  denote $\E(f(Z))=\int f(z)\dd\P(z)$ in the main manuscript). In
  this proof we use the alternative notation
  $\D_\eta(Z)=\D_1(Z;\eta)$. By definition of the TMLE and the fact
  that it solves the efficient influence function estimating equation,
  we have
  \[\hat \U_1 = \frac{1}{J}\sum_{j=1}^J\Pnj \D_{\tilde \eta_j}.\]
  Thus,
  \begin{equation}
    \sqrt{n}(\hat \U_1 - \U_1)=\Gn (\D_\eta
    - \U_1) + R_{n,1} + R_{n,2},\label{eq:exptmle}
  \end{equation}
  where
  \[  R_{n,1}  =\frac{1}{\sqrt{J}}\sum_{j=1}^J\Gnj(\D_{\tilde
      \eta_j} - \D_{\eta}),\,\,\,
    R_{n,2}  = \frac{\sqrt{n}}{J}\sum_{j=1}^J\P(\D_{\tilde
      \eta_j}-\theta).
  \]
  Theorem~\ref{lemma:von} together Lemma~\ref{lemma:remorder} and
  the assumptions of the theorem shows that $R_{n,2}=o_\P(1)$.

  Let $F_n^j=\D_{\hat \eta_j,\epsilon} - \D_\eta$ and $\mathcal F_n^j$
  denote the class with one element equal to $F_n^j$. Because the
  function $\hat\eta_j$ is fixed given the training data, we can apply
  Theorem 2.14.2 of \cite{vanderVaart&Wellner96} to obtain
  \begin{equation}
    E\left\{|\Gnj F_n^j| \,\,\bigg|\,\, {\cal
        T}_j\right\}\lesssim \lVert F^j_n \rVert \int_0^1
    \sqrt{1+N_{[\,]}(\alpha \lVert F_n^j \rVert, {\cal F}_n^j,
      L_2(\P))}\dd\alpha,\label{eq:empproc}
  \end{equation} where
  $N_{[\,]}(\alpha \lVert F_n^j \rVert, {\cal F}_n^j, L_2(\P))$ is the
  bracketing number. Theorem 2.7.2 of \cite{vanderVaart&Wellner96} shows
  \[\log N_{[\,]}(\alpha \lVert F_n^j \rVert, {\cal F}_n^j, L_2(\P))\lesssim
    \frac{1}{\alpha \lVert F_n^j \rVert}.\]
  This shows
  \begin{align*}
    \lVert F^j_n \rVert \int_0^1\sqrt{1+N_{[\,]}(\alpha
    \lVert F_n^j \rVert, {\cal F}_n^j, L_2(\P))}\dd\alpha &\lesssim
                                                            \int_0^1\sqrt{\lVert F^j_n \rVert^2+\frac{\lVert F^j_n
                                                            \rVert}{\alpha}}\dd\alpha\\ &\leq
                                                                                          \lVert F^j_n \rVert + \lVert F^j_n \rVert^{1/2}
                                                                                          \int_0^1\frac{1}{\alpha^{1/2}}\dd\alpha\\
                                                          &\leq \lVert F^j_n \rVert + 2 \lVert F^j_n \rVert^{1/2}.
  \end{align*}
  Since $\hat\eta$ is consistent and $\delta_n\to 0$, $\lVert F^j_n \rVert = o_P(1)$. This
  shows $\sup_{f\in {\cal F}_n^j}\Gnj f=o_P(1)$ for each $j$,
  conditional on ${\cal T}_j$. Thus $
  R_{n,1}=o_\P(1)$.

  This shows
  \[\hat \U_1 -\U_1 = \frac{1}{n}\sum_{i=1}^n[\D_1(Z_i;\eta) -\U_1] +
    o_\P(n^{-1/2}).\]
  By definition of $\hat\beta_{\text{\scriptsize sdr}}$, we have
  \[\U_2(\hat\beta_{\text{\scriptsize sdr}}) + \hat \U_1 =
    o_\P(n^{-1/2}),\]
  which yields
  \[\U_2(\hat\beta_{\text{\scriptsize sdr}}) + \frac{1}{n}\sum_{i=1}^n\D_1(Z_i;\eta) =
    o_\P(n^{-1/2}).\] Applying standard theory for M-estimation
  \citep[e.g., Theorem 5.23 of][]{vanderVaart98} and noticing that
  $\U_2(\beta) = -\U_1$ yields the desired result.
\end{proof}

\section{TMLE-like Estimator \label{tmle-est}}

Our proposed targeted maximum likelihood estimator (TMLE) is a simple naive extension of the original TMLE as presented in \citep{GruberLaan2009}. 

Define the nuisance parameter $\eta_t = \left( \r_t, T_t \right)$ defined as:

\begin{align*}
    \r_t(a_t, h_t) & =\frac{\lambda_t(a_t\mid \bar a_{t-1}, v)}{g_t(a_t\mid h_t)} \\
    \T_t:(a_t, h_t)&\mapsto  \E[\bar \T_{t+1}(H_{t+1})\mid
                     A_t=a_t, H_t = h_t]
\end{align*}

And implemented the following algorithm:

\begin{enumerate}[label=Step \arabic*, align=left, leftmargin=*]
\item Initialize
  $\check{\bar T}_{\tau + 1,i}^* = Y_i \times \phi(\bar A,V)$
  for $i=1,\ldots,n$.
\item For $t=\tau,\ldots,1$:
  \begin{enumerate}[label=(\roman*)]
  \item Compute the pseudo-outcome
    $\check Y_{t+1,i} =
    \check{\bar T}_{t + 1,i}^*$ for all
    $i=1,\ldots,n$.
  \item For $j=1,\ldots,J$:
    \begin{itemize}
    \item Regress $\check Y_{t+1,i}$ on $(A_{t,i}, H_{t,i})$ using any
      regression technique and using only data points
      $i\in \mathcal T_{j}$.\label{step:TMLE2}
    \item Let $\check T_{t,j}$ denote the output, update
      $\check\eta_{t, j} = (\hat r_{t,j}, \check
      T_{t,j},\ldots,\hat r_{\tau,j}, \check T_{\tau,j})$, and
      iterate.
    \item Compute $\check{\bar T}_{t,j}$ by numerical integration or
      importance sampling
    \end{itemize}
    \item Compute $\check{\bar T}_{t}^* = \check{\bar T}_{t} + \hat \epsilon$, where $\hat \epsilon$ is the solution to the estimating equation \\
    $n^{-1} \sum_{i=1}^n \left( \prod_{s = t}^{\tau} \hat r_{s,i} \right) \left[  
\check{ \bar T}_{t+1,i} - \left( \check T_{t+1,i} + \hat{\epsilon} \right) \right] = 0$. We get an ordinary least square estimate of $\hat \epsilon$ with an intercept only regression of $\check Y_{t+1,i}$ on offset $\check T_{t,i}$ weighted by $\prod_{s = t}^{\tau} \hat r_{s,i}$.
  \end{enumerate}
\item Define $\hat \U_1=n^{-1}\sum_{i=1}^n \check{\bar T}_i^*$.
\item Solve $\hat \U_1+\U_2(\gamma)=0$:
  \begin{enumerate}[label=(\roman*)]
  \item Initialize $k=0$ and $\hat \beta_k = \hat\beta_{\text{\scriptsize ipw}}$;
  \item Let $\hat \beta_{k+1} = \hat\beta_k - [\dot
    \U_2(\hat\beta_k)]^{-1}[\hat \U_1+\U_2(\hat\beta_k)]$;\label{stepTMLE1}
  \item Update $k=k+1$\label{stepTMLE2};
  \item Iterate \ref{stepTMLE1} and \ref{stepTMLE2} until convergence, i.e.,
    until $\U_2(\hat\beta_k)=\hat \U_1 + o_\P(n^{-1/2})$.
  \item Let
    $\hat\beta_{\text{\scriptsize TMLE}}$ denote the resulting estimator.
  \end{enumerate}
\end{enumerate}

\section{Simulation Results} \label{appsec:simres}

\begin{table}[ht]

\caption{Simulation results summary.}
\centering
\small 
\begin{tabular}[t]{r|r|r|r|r|r|r|r|r|r|r|r}
\hline
\multicolumn{1}{c|}{N} & \multicolumn{3}{c|}{Bias} & \multicolumn{3}{c|}{ $\sqrt{n} \times $ Bias} & \multicolumn{3}{c|}{$n \times$ MSE} & \multicolumn{2}{c}{Coverage} \\
\cline{1-1} \cline{2-4} \cline{5-7} \cline{8-10} \cline{11-12}
 & SDR & TMLE & IPW & SDR & TMLE & IPW & SDR & TMLE & IPW & SDR & TMLE\\
\hline
\multicolumn{12}{l}{\textbf{Scenario 1}} \\
\hline
250 & 0.03 & 0.04 & 0.06 & 0.52 & 0.63 & 0.93 & 1.15 & 1.24 & 1.78 & 0.84 & 0.83 \\ 
\hline
  500 & 0.01 & 0.01 & 0.02 & 0.15 & 0.15 & 0.54 & 0.69 & 0.69 & 1.04 & 0.92 & 0.94 \\ 
  \hline
  1000 & 0.00 & 0.00 & 0.02 & 0.09 & 0.07 & 0.58 & 0.50 & 0.52 & 1.00 & 0.96 & 0.95 \\ 
  \hline
  2000 & 0.00 & 0.00 & 0.01 & 0.02 & 0.00 & 0.48 & 0.53 & 0.53 & 0.89 & 0.95 & 0.95 \\ 
  \hline
  3000 & 0.00 & 0.00 & 0.01 & 0.07 & 0.06 & 0.42 & 0.50 & 0.50 & 0.75 & 0.94 & 0.94 \\ 
  \hline
  4000 & 0.00 & 0.00 & 0.01 & 0.10 & 0.10 & 0.34 & 0.56 & 0.54 & 0.71 & 0.94 & 0.94 \\ 
  \hline
  5000 & 0.00 & 0.00 & 0.00 & 0.01 & 0.01 & 0.13 & 0.54 & 0.53 & 0.53 & 0.94 & 0.94 \\
\hline
\multicolumn{12}{l}{\textbf{Scenario 2}}\\
\hline
250 & 0.06 & 0.08 & 0.16 & 0.91 & 1.31 & 2.51 & 2.05 & 2.75 & 7.10 & 0.70 & 0.56 \\ 
\hline
  500 & 0.02 & 0.03 & 0.15 & 0.54 & 0.75 & 3.38 & 1.19 & 1.42 & 12.26 & 0.84 & 0.76 \\ 
  \hline
  1000 & 0.01 & 0.01 & 0.15 & 0.32 & 0.39 & 4.88 & 0.91 & 0.88 & 24.59 & 0.88 & 0.86 \\ 
  \hline
  2000 & 0.00 & 0.00 & 0.15 & 0.08 & 0.19 & 6.82 & 0.66 & 0.68 & 47.18 & 0.94 & 0.94 \\ 
  \hline
  3000 & 0.00 & 0.00 & 0.16 & 0.06 & 0.16 & 8.52 & 0.79 & 0.72 & 73.25 & 0.88 & 0.92 \\ 
  \hline
  4000 & 0.00 & 0.00 & 0.15 & 0.04 & 0.15 & 9.80 & 0.94 & 0.71 & 96.76 & 0.86 & 0.92 \\ 
  \hline
  5000 & -0.00 & 0.00 & 0.15 & -0.14 & 0.03 & 10.77 & 0.86 & 0.64 & 116.60 & 0.89 & 0.95 \\ 
\hline
\multicolumn{12}{l}{\textbf{Scenario 3}}\\
\hline
  250 & 0.08 & 0.08 & 0.06 & 1.25 & 1.20 & 0.93 & 2.95 & 2.84 & 1.78 & 0.90 & 0.91 \\ 
  \hline
  500 & 0.04 & 0.04 & 0.02 & 0.89 & 0.79 & 0.54 & 2.37 & 2.26 & 1.04 & 0.97 & 0.97 \\
  \hline
  1000 & 0.03 & 0.03 & 0.02 & 1.01 & 0.89 & 0.58 & 2.50 & 2.27 & 1.00 & 0.97 & 0.98 \\
  \hline
  2000 & 0.02 & 0.02 & 0.01 & 0.83 & 0.74 & 0.48 & 2.07 & 1.88 & 0.89 & 0.98 & 0.98 \\
  \hline
  3000 & 0.01 & 0.01 & 0.01 & 0.63 & 0.57 & 0.42 & 1.72 & 1.61 & 0.75 & 0.99 & 0.99 \\
  \hline
  4000 & 0.00 & 0.00 & 0.01 & 0.31 & 0.28 & 0.34 & 1.47 & 1.42 & 0.71 & 1.00 & 1.00 \\
  \hline
  5000 & 0.00 & 0.00 & 0.00 & 0.03 & 0.02 & 0.13 & 1.23 & 1.20 & 0.53 & 1.00 & 1.00 \\
\hline
\multicolumn{12}{l}{\textbf{Scenario 4}}\\
\hline
  250 & 0.08 & 0.13 & 0.12 & 1.26 & 2.01 & 1.83 & 2.87 & 5.15 & 4.25 & 0.85 & 0.77 \\ 
  \hline
  500 & 0.04 & 0.10 & 0.09 & 0.93 & 2.24 & 1.99 & 2.13 & 5.94 & 4.57 & 0.93 & 0.77 \\ 
  \hline
  1000 & 0.03 & 0.10 & 0.09 & 0.91 & 3.10 & 2.75 & 2.05 & 10.51 & 8.20 & 0.94 & 0.49 \\ 
  \hline
  2000 & 0.02 & 0.09 & 0.08 & 0.73 & 4.08 & 3.70 & 1.74 & 17.58 & 14.39 & 0.96 & 0.22 \\ 
  \hline
  3000 & 0.01 & 0.09 & 0.08 & 0.64 & 4.70 & 4.44 & 1.73 & 23.25 & 20.46 & 0.96 & 0.13 \\ 
  \hline
  4000 & 0.01 & 0.08 & 0.08 & 0.55 & 5.20 & 5.00 & 1.60 & 27.93 & 25.58 & 0.98 & 0.03 \\ 
  \hline
  5000 & 0.00 & 0.08 & 0.08 & 0.24 & 5.47 & 5.33 & 1.31 & 30.65 & 28.96 & 1.00 & 0.01 \\
\hline
\multicolumn{12}{l}{\textbf{Scenario 5}}\\
\hline
  250 & 0.06 & 0.07 & 0.10 & 0.99 & 1.09 & 1.59 & 1.94 & 2.07 & 3.36 & 0.90 & 0.91 \\ 
  \hline
  500 & 0.03 & 0.04 & 0.09 & 0.75 & 0.83 & 1.93 & 1.49 & 1.55 & 4.65 & 0.96 & 0.94 \\ 
  \hline
  1000 & 0.02 & 0.02 & 0.09 & 0.60 & 0.64 & 2.73 & 1.16 & 1.13 & 8.22 & 0.96 & 0.98 \\ 
  \hline
  2000 & 0.00 & 0.01 & 0.08 & 0.19 & 0.27 & 3.59 & 0.69 & 0.72 & 13.64 & 1.00 & 1.00 \\ 
  \hline
  3000 & 0.00 & 0.00 & 0.08 & 0.02 & 0.09 & 4.43 & 0.76 & 0.67 & 20.36 & 1.00 & 1.00 \\ 
  \hline
  4000 & -0.00 & -0.00 & 0.08 & -0.17 & -0.10 & 5.06 & 0.85 & 0.71 & 26.38 & 1.00 & 1.00 \\ 
  \hline
  5000 & -0.01 & -0.00 & 0.08 & -0.47 & -0.31 & 5.50 & 0.92 & 0.70 & 30.84 & 1.00 & 1.00 \\ 
\hline
\end{tabular}
\label{tab:tab3}
\end{table}
